\pdfoutput=1
\documentclass[twocolumn,twocolappendix]{aastex63}

\usepackage{natbib,color,graphicx}
\usepackage{textcomp}
\usepackage{float}
\usepackage{soul}
\usepackage{enumerate}
\usepackage{amsmath}

\accepted{October 11, 2021}
\submitjournal{ApJ}

\begin{document}

\title{Shock-wave radio probing of solar wind sources in coronal magnetic fields}

\correspondingauthor{Artem Koval}
\email{koval@asu.cas.cz}

\author{Artem Koval}
\affiliation{Astronomical Institute, Czech Academy of Sciences, 251 65 Ond\v{r}ejov, Czech Republic}

\author{Marian Karlick$\acute{\textrm{y}}$}
\affiliation{Astronomical Institute, Czech Academy of Sciences, 251 65 Ond\v{r}ejov, Czech Republic}

\author{Aleksander Stanislavsky}
\affiliation{Institute of Radio Astronomy, National Academy of Sciences of Ukraine, 61002 Kharkiv, Ukraine;\\
Faculty of Pure and Applied Mathematics, Hugo Steinhaus Center,\\ Wroc{\l}aw University of Science and Technology, 50-370 Wroc{\l}aw, Poland}

\author{Bing Wang}
\affiliation{Shandong Provincial Key Laboratory of Optical Astronomy and Solar-Terrestrial Environment, \\ School of Space Science and Physics, Institute of Space Sciences, Shandong University, 264209 Weihai, China}

\author{Miroslav B$\acute{\textrm{a}}$rta}
\affiliation{Astronomical Institute, Czech Academy of Sciences, 251 65 Ond\v{r}ejov, Czech Republic}

\author{Roman Gorgutsa}
\affiliation{Pushkov Institute of Terrestrial Magnetism, Ionosphere and Radio Wave Propagation, \\ Russian Academy of Sciences, 108840 Troitsk, Russia}

\begin{abstract}

The Space Weather effects in the near-Earth environment as well as in atmospheres of other terrestrial planets arise by corpuscular radiation from the Sun, known as the solar wind.  The solar magnetic fields govern the solar corona structure. Magnetic-field strength values in the solar wind sources -- key information for modeling and forecasting the Space Weather climate -- are derived from various solar space- and ground-based observations, but, so far not accounting for specific types of radio bursts. These are ``fractured'' type II radio bursts attributed to collisions of shock waves with coronal structures emitting the solar wind. Here, we report about radio observations of two ``fractured'' type II bursts to demonstrate a novel tool for probing of magnetic field variations in the solar wind sources. These results have direct impact on interpretations of this class of bursts and contribute to the current studies of the solar wind emitters.

\end{abstract}


\section{Introduction}\label{sec:intro}

The continuous outflow of charged particles rushing from the solar corona -- known as \textit{solar wind} -- is of particular interest in the study of space plasmas and heliophysics~\citep{Parker1958}. The heliosphere, being a bubble-like cavity in the interstellar medium, is permeated by the solar wind plasma dragging the solar magnetic field throughout in the Solar System~\citep{Dessler1967}. The knowledge about the origin and properties of the solar wind obtained from various solar observations is of crucial importance for modeling and forecasting the Space Weather effects in the near-Earth space environment. They can cause failures/disruptions in space- and ground-based technological systems as well as to impair human's health or even endanger his life~\citep{Blanch2013}. The Space Weather climate of other terrestrial planets, like Venus and Mars, currently is highly popular subject of scientific researches~\citep{Collinson2012, Hassler2018}. If for the former our exploration remains purely basic-oriented, for the latter it has become thoroughly scrutinizing in connection with ahead-of-its-time plans of martian colonization to support and ensure potential space human missions there in the nearest future~\citep{Musk2017}.

Parker\textquotesingle s theoretical work is esteemed to be a milestone in studying solar wind sources~\citep{Parker1958}, the contribution of which after exactly 60 years has been noticed in launch of the Parker Solar Probe spacecraft~\citep{Bale2019}. It has been long recognized, there are fast ($\simeq$750 km s$^{-1}$) and slow ($\simeq$400 km s$^{-1}$) flows of the solar wind discriminated by their velocities as measured at the Earth orbit~\citep{Feldman2005}. While the solar coronal holes are firmly established as sources of the fast flow~\citep{Krieger1973}, the sources of the slow flow are still under debates~\citep{Kilpua2016}. The magnetic field of the Sun governs the structure of the solar corona where the solar wind emanates and further accelerates supersonically. Therefore, the accurate observational data about the topology, and more significantly, quantities of the coronal magnetic field are pivotal for identifying the solar wind sources as well as for the Space Weather modeling where these data specify initial conditions.

To date, the radio observations are considered as ones of the most reliable methods for obtaining the absolute magnetic-field strength in the solar corona~\citep{Alissandrakis2021}. A particular class of solar radio bursts -- type~II bursts -- is used to probe ambient coronal magnetic fields~\citep{Vrsnak2002}. On solar spectrograms type II bursts usually drift gradually to lower frequencies as one or two bright emission stripes, with close to a 2:1 frequency ratio, corresponding to emissions at the fundamental and harmonic plasma frequencies. It is generally accepted that their origin is shock waves, propagating in the solar corona and driven by powerful eruptive phenomena (i.e., flares or/and coronal mass ejections (CMEs))~\citep{Mann1996,Morosan2019}. Thus, the shock wave accelerates ahead-of-it electrons, generating Langmuir or plasma waves which then are converted into the electromagnetic radio emission. The sources of the type II radio emission locate above the nose and/or the flanks of the CME-driven shock~\citep{Mancuso2004,Makela2012}.

Occasionally, these bursts demonstrate distinctive spectral morphology. Particularly, the emission stripe may have the band-splitting structure whereby each stripe splits into two bands. The most widely accepted model of the band-splitting explains this spectral feature by emission from two electron density regions ahead (upstream) and behind (downstream) the shock front~\citep{Smerd1975}. Recent relevant studies on type II radio busts with the band-splitting observed by means of the LOw-Frequency ARray (LOFAR; \citet{vanHaarlem2013}) support this model~\citep{Chrysaphi2018,Chrysaphi2020}. In particular, \citet{Chrysaphi2018} investigated a spatial separation between the upper and lower band type II radio sources on imaging measurements with LOFAR. The authors first have quantitatively described such separation considering radio-wave propagation effects in the solar corona, mainly scattering which is the dominant for meter-decameter radio waves~\citep{Kontar2017}. These important results indicate that the upper and lower band radio emission sources have nearly cospatial locations. In the framework of the band-splitting model, the spectral pattern analysis allows ones to provide diagnostics of magnetic fields in the solar corona~\citep{Vrsnak2002,Stanislavsky2015}.

Furthermore, at times type II bursts exhibit spectrally indented shapes in the form of bumps or breaks. These ``fractured'' radio busts are the most spectacular and unexplored yet. They are attributed to the collisions of shock waves with coronal structures (so far only streamers), being solar wind sources~\citep{Kong2012,Feng2013}. Here, we should also note the study by \citet{Gao2016} reporting the broken spectral structure of a type II radio burst caused by interaction of a coronal shock with a flare current sheet.

Despite the plenty of works on the routine analysis of ordinary (non-indented) type II events relative to estimating magnetic fields in the corona, there have been no measurements of the magnetic field in the solar wind sources by examining ``fractured'' type II bursts that can be set up as a new precise technique. In this study, we analyze radio observations of two successive type II bursts which are ``fractured'' and band-splitted at the same time. By the example of careful investigations of these bursts, for which the magnetic field variations in a pseudo-streamer and flux tube have been derived, we deploy up-to-date robust tools for probing solar wind sources.

\begin{figure*}[!t]
\begin{center}
\includegraphics[width=0.7\textwidth]{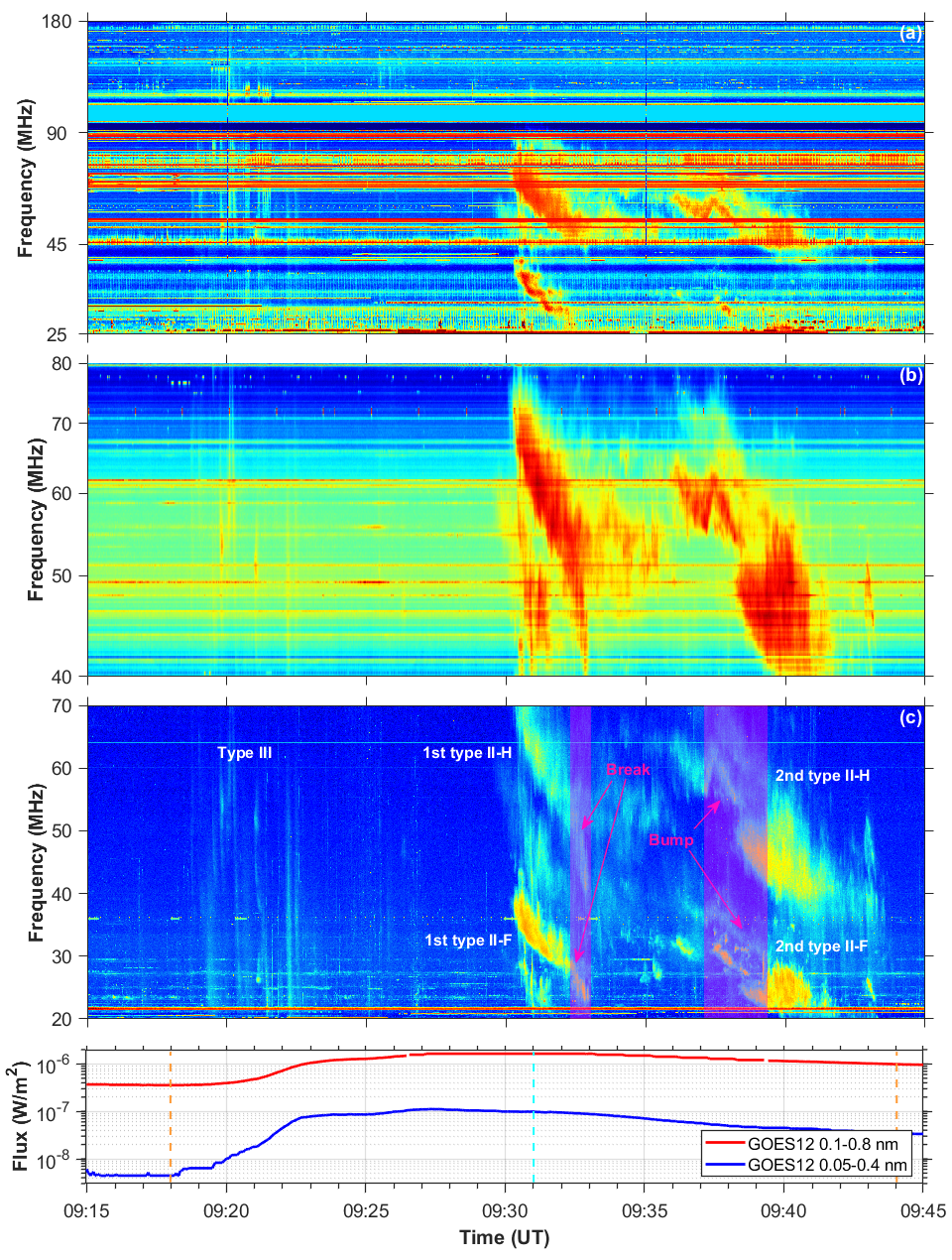}
\caption{Collection of the time-aligned solar dynamic spectra with the IZMIRAN radio observatory (25--180 MHz; Troitsk, Russia) (a), the AIP radio spectropolarimeter (40--80 MHz; Tremsdorf, Germany) (b), the Nan\c{c}ay Decametric Array (20--70 MHz; Nan\c{c}ay, France) (c) on March 17th, 2004. The fundamental (F) and harmonic (H) components of fractured type II bursts are labeled on the NDA spectrogram, where the semitransparent regions indicate approximate durations of break and bump spectral features. A group of type III bursts is denoted. The lower panel is the solar X-ray emission measured by GOES 12 on that day. The C1.6 class flare start/end and peak times are pointed by orange and cyan dashed vertical lines, respectively. The common time axis is in the range 09:15--09:45 UT.}
\label{Figure1}
\end{center}
\end{figure*}

\section{Results}

\subsection{Overview of the multi-band observations}

\begin{figure*}[!t]
\begin{center}
\includegraphics[width=0.99\textwidth]{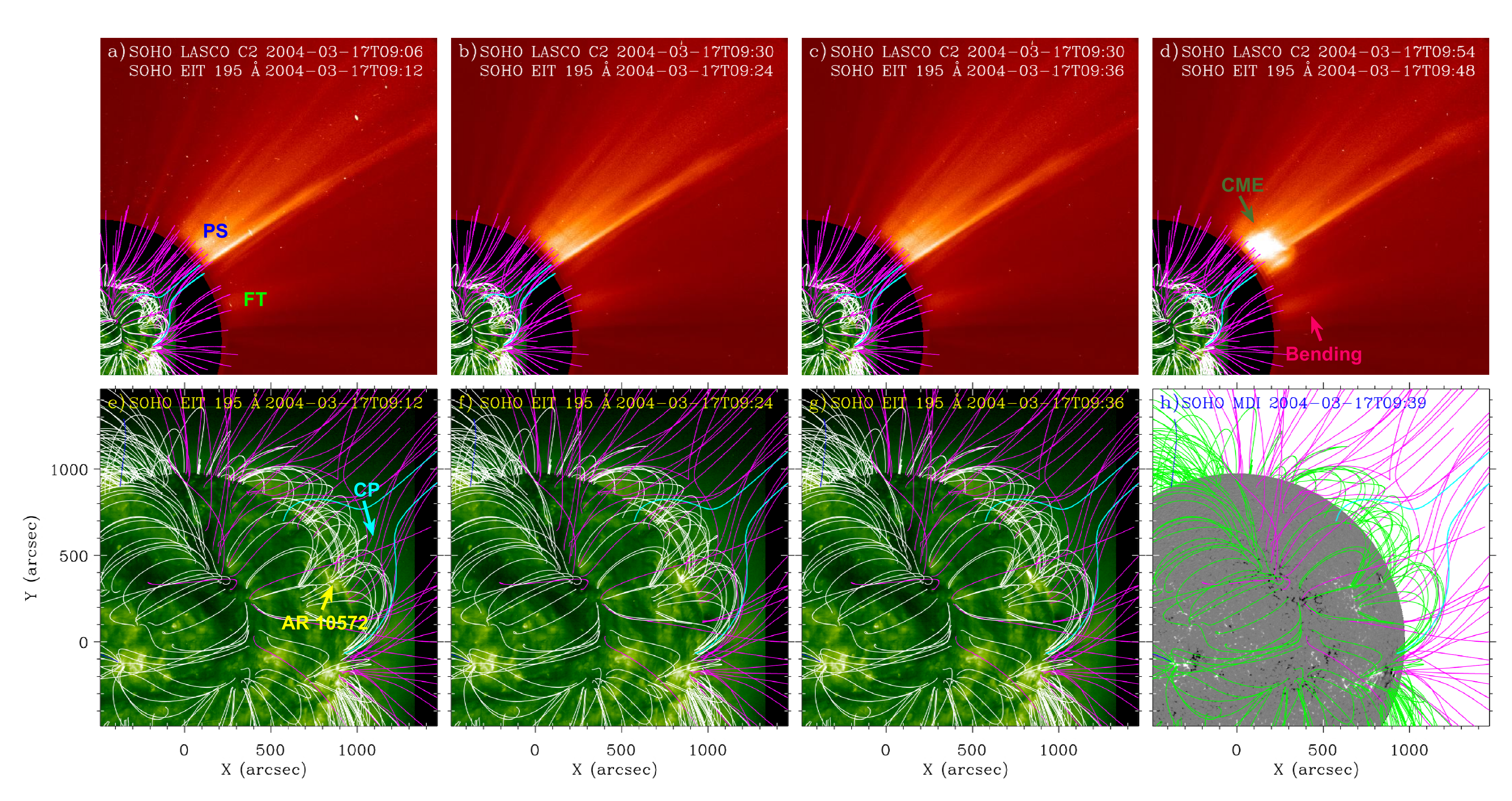}
\caption{Magnetic field configurations obtained from the PFSS model based on the measurements from the SOHO MDI for the Carrington Rotation 2014. In panels (a-g) field lines are superimposed on SOHO LASCO C2 and SOHO EIT 195 {\AA} images, where the closed field lines are colored in white, while inward (outward) field lines are pictured in purple (blue), respectively. In panel (h) the SOHO MDI magnetogram at 09:39 UT shows the magnetic regions of the south (dark color) and north (white color) polarity on the photosphere, where the closed field lines are colored in green, while colors of outward (inward) field lines correspond to the ones in panels (a-g). In panel (a) labels identify structures as the pseudo-streamer (PS) and the flux tube (FT). The site of the NOAA active region (AR) 10572 is pointed by the yellow arrow in panel (e). There an approximate location of a cusp point of the PS is indicated by the cyan arrow. In panel (d) the pink arrow denotes a visible bending of the FT caused by the CME passage; compare with panel (c).}
\label{Figure2}
\end{center}
\end{figure*}

On 17 March 2004, the moderate C1.6 class solar flare was detected by the Geostationary Operational Environmental Satellite 12 (GOES-12; \citep{Neupert2011}) measuring the solar X-ray flux in two wavelength bands, 0.5--3.0 {\AA} and 1.0--8.0 {\AA}. The flare erupted from the NOAA active region (AR) 10572 (N21W66 at 09:20 UT) having the $\alpha$ class within the ARs' magnetic classification that manifests its unipolarity. The flare was connected with a relatively narrow and medium-speed CME, showing the angular width of 53$^\circ$ and the velocity of 718 km~s$^{-1}$, observed at C2 white-light coronagraph images from the Large Angle Spectroscopic Coronagraph (LASCO) aboard the Solar and Heliospheric Observatory (SOHO) spacecraft~\citep{Brueckner1995}. The field of view (FOV) and the cadence of the LASCO/C2 were 2.2--6 solar radii and 24 minutes, respectively. The imaging observations were supplemented by the Extreme ultraviolet Imaging Telescope (EIT) data from the SOHO mission as well~\citep{Delaboudiniere1995}. This instrument provides images of the Sun, particularly, at 195 {\AA} up to 1.5 solar radii with a 12-minute cadence.

The flare and CME were associated with solar bursts activity in the form of types III and II bursts recorded in meter-decameter radio observations by the IZMIRAN radio telescopes (25--180 MHz; time and frequency resolutions: 1 s and 140 kHz (25--45 MHz), 250 kHz (45--90 MHz), 500 kHz (90--180 MHz); Troitsk, Russia; \citep{Gorgutsa2001}), the AIP radio spectropolarimeter (40--80 MHz; time and frequency resolutions: 0.1 s and 235 kHz; Tremsdorf, Germany; \citep{Mann1992}), the Nan\c{c}ay Decametric Array (NDA) (20--70 MHz; time and frequency resolutions: 1 s and 125 kHz; Nan\c{c}ay, France; \citep{Lecacheux2000})(Figure~\ref{Figure1}). A group of type III radio bursts appeared on the dynamic spectra shortly after the flare onset at $\sim$09:18 UT. This type of radio bursts is a signature of accelerations of electron beams in a flare site, which then propagate typically at about one third of the light speed in the solar atmosphere~\citep{Reid2014}. Type III bursts often forebode subsequent occurrence of type II radio bursts in dynamic spectra, as in the present case. Figure~\ref{Figure1} shows two prominent type II bursts after the multiple type III bursts. Note that they had ``fractured'' spectral shapes. The spectrograms registered by means of three radio instruments, located in different places in Earth, exhibit spectral similarity of the corresponding bursts. Thus, it entirely excludes possible instrumental or ionospheric effects to which unusual ``fracturness'' of type II bursts may be imputed.

The two successive type II bursts demonstrate a collection of important spectral features. While the first type II burst has the spectral break, the second type II burst has the spectral bump, and moreover both type II bursts show the band-splitting. The close in time appearance of the ``fractured'' type II events are firmly linked to the CME motion in the solar corona, since there are no other observed eruptions. This implies the only single shock, produced by the CME, traveled through different coronal structures that is manifested in the indented spectral morphology of both bursts. The event is outstanding on two grounds. First, an appearance of the type II bursts with the break and bump in close timeline is extraordinary. Second, the band-splitting patterns in both bursts give an opportunity to estimate the magnetic-field strength in quasi-stationary coronal structures belonging to solar wind emitters. This is a novel approach in studying the sources of the solar wind.

\subsection{Imaging observations}

The SOHO observations offer the possibility of imaging the Sun in EUV and white light. We have treated these data to trace the CME/shock propagation and identify quasi-stationary plasma structures in the solar corona. Besides, the modeling of the magnetic field lines in the solar corona has been applied. Figure~\ref{Figure2}(a-d) shows the fusing of the observational data from the LASCO/C2 and EIT, as well as the magnetic field lines topology obtained by means of the Potential-Field Source-Surface (PFSS) model~\citep{Wang1992}. Measurements of magnetic fields at the photosphere were made by the Michelson Doppler Imager (MDI) aboard the SOHO. By the extrapolation of photospheric magnetic fields with the PFSS model, the global magnetic field up to 2.5 solar radii was reproduced. Figure~\ref{Figure2}(e-g) corresponds to Figure~\ref{Figure2}(a-c) with no coronagraphic LASCO/C2 images, thus demonstrating a closer view. Figure~\ref{Figure2}(h) shows the magnetogram at $\sim$09:39 UT.

In the imaging study, two coronal structures are in focus, namely a pseudo-streamer (PS) and a flux tube (FT). The PS is discernible in Figure~\ref{Figure2}(a-d). Its distinctive configuration is delineated by open field lines (inward) of the same (south) polarity. The inward open field lines (purple color), rising up from the photosphere, firstly converge in a cusp point (CP), thus forming a semi-circular closed field region below it, and then stretch almost radially. The PS is elongated in the direction away from an observer and oriented at an angle to the sky plane, so that on the LASCO/C2 image the PS rays look like a wide bright fan. The approximate height of the low-lying CP is about 1.5 solar radii. The semi-circular region below the CP contains two loop arcades portrayed by the closed field lines (white color). The loops, whose opposite footpoints attach to south polarity regions, are rooted in the unipolar AR 10572 by their adjacent footpoints. The AR has north polarity and locates directly below the CP. The foresaid description suits for PSs which have been recently categorized as a new kind of white-light streamers as well as being suppliers of the hybrid-speed solar wind in the heliosphere~\citep{Wang2007,Wang2012}.

The FT in Figure~\ref{Figure2}(a-d) is recognizable by its conical shape formed by a bundle of open field lines. There the inward open field lines emerge up from the photosphere and rapidly expand in the corona. In the LASCO/C2 images the expanded FT covers several tens of degrees in longitude at distance of 2.2 solar radii. It is seen that one footpoint of the FT is rooted in the same parent region of the south polarity to which one side of the PS is attached. Thus, the PS and FT are in contact with each other. The remote footpoint of the FT extends outward beyond several solar radii in the coronagraphic image, that allows ones to consider the FT ``open'' to the heliosphere. In this way, FTs provide a transport of solar wind particles from localized acceleration sites in an active region to the heliosphere~\citep{Pick2006,Klein2008}.

Figure~\ref{Figure2}(a,e) represents the pre-erupted state preceding the C1.6 flare and CME. Figure~\ref{Figure2}(b,c,f,g) corresponds to the active phase of the flare. Though there are enhancements in the AR, indicating the process of energy release, the CME propagation is not recognizable in two successive EIT images (Figure~\ref{Figure2}(f,g)). The low-cadence coronagraphic measurements do not provide the detailed evolutional track of the CME as well. Thus, at 09:30~UT, close to the beginning of the first type II burst, the CME was obscured by the occulting disk, according to the LASCO/C2 (Figure~\ref{Figure2}(b,c)). The CME appeared in the subsequent observation at 09:54~UT (Figure~\ref{Figure2}(d)), $\sim$10 minutes later the second type II burst terminated. Nevertheless, there is an obvious collision of the PS and CME. The latter penetrated into the PS's structure during its rising. Besides, by examining the coronagraphic images in Figure~\ref{Figure2}(c,d), the FT demonstrates a visible bending in Figure~\ref{Figure2}(d), which is not seen in the previous image. It reveals that during its expansion in the corona the CME bounced the FT aside.

The careful analysis of a synergy of imaging and spectral measurements has permitted us to state that: i) the CME was the only observed disturbance, which is responsible for the shock wave; ii) both ``fractured'' type II emissions occupy approximately the same frequency range and appear with the interval of 3 minutes; iii) the only coronal structures, which laid in the same spatial quadrant and interacted with the CME/shock, were the pseudo-streamer and flux tube. From these studies, we infer that the type II bursts with those specific spectral shapes could only appear in a sequence if their radio emission sources located at nose and flank of the same shock wave. Hence, the radio source of the type II burst, having spectral break, was placed at the shock nose, and the spectral shape is resulted from the radio source movement through the PS's structure from inside to out. The radio source of the type II burst with spectral bump was situated at the shock flank, whereas the spectral bump is caused by the propagation of its radio source across the FT.

\subsection{``Fractured'' type II burst with spectral break}

\begin{figure}[!t]
\begin{center}
\includegraphics[width=1.0\hsize]{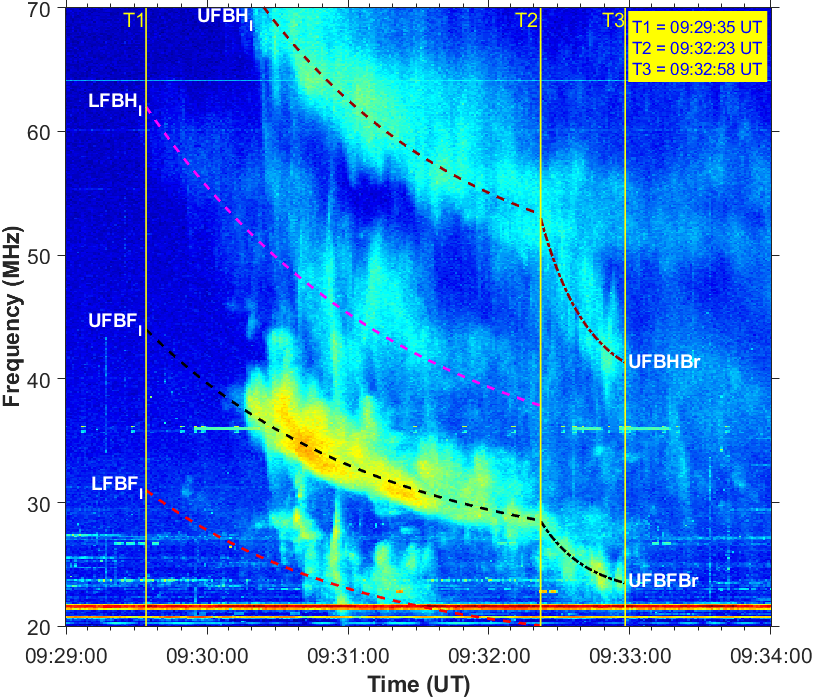}
\caption{The NDA dynamic spectrum of the type II radio burst with spectral break. Lower/upper frequency bands of fundamental and harmonic components are labeled as LFBF$_{\textrm{I}}$/UFBF$_{\textrm{I}}$ and LFBH$_{\textrm{I}}$/UFBH$_{\textrm{I}}$, respectively. The post-break bands presenting in UFBF$_{\textrm{I}}$ and UFBH$_{\textrm{I}}$ are designated as UFBFBr and UFBHBr, respectively. Fitting with the exponential function is performed for all bands. The yellow vertical lines indicate time instants T1, T2, and T3.}
\label{Figure3}
\end{center}
\end{figure}

Figure~\ref{Figure3} shows the NDA spectrogram with the first ``fractured'' type II burst. The radio emission onset is at 09:29:35 UT (T1). It demonstrates both harmonic structure and band-splitting. The fundamental (F) and harmonic (H) components split on lower and upper frequency bands, thereby dubbed as LFBF$_{\textrm{I}}$ and UFBF$_{\textrm{I}}$, and LFBH$_{\textrm{I}}$ and UFBH$_{\textrm{I}}$, respectively. The most important spectral feature of this type II burst is the spectral break recognizable only in the UFBF$_{\textrm{I}}$ and UFBH$_{\textrm{I}}$. The break point is at 09:32:23 UT (T2). After that instant the UFBF$_{\textrm{I}}$ and UFBH$_{\textrm{I}}$ do not continue their gradual dropping, but undergo abrupt falling that visually fractures the bands on pre-break and post-break parts. The post-break emissions, designated as UFBFBr and UFBHBr, ended at 09:32:58 UT (T3), in 35 seconds after the break point. This spectral morphology of a type II burst is associated with the interaction of a shock wave and a streamer, namely when a type II emission source pierces the streamer from inside to out.

From the imaging observations we have identified the CME/shock interacting with the PS. The pre-break radio emission of the type II burst relates to the shock wave propagation in the closed magnetic region whose apex is the CP. The pre-break band-splitting of the type II burst can be used to infer the magnetic field in the lower part of the PS, namely below its CP. The expression for the magnetic-field strength there, $B_{\textrm{PS}}$, takes the form:
\begin{equation}
     B_{\textrm{PS}}[\textrm{G}] = 5.1\cdot10^5\cdot f_{\textrm{LFBF}_{\textrm{I}}}[\textrm{MHz}]\cdot V_{S_{N}}[\textrm{km}~\textrm{s}^{-1}]/M_A,
     \label{Eq1}
\end{equation}
where $f_{\textrm{LFBF}_{\textrm{I}}}$, $M_A$, and $V_{S_{N}}$ values were found (see Appendices~\ref{appendixA}(Table~\ref{Table1}), \ref{appendixB}, and \ref{appendixC}).

Let us here focus on the $V_{S_{N}}$ parameter, which is the speed of the shock wave at its nose. Because of a lack of direct time-distance observations of the shock wave in the corona, the shock speed was calculated using the frequency drift rate (FDR), $df_{\textrm{LFBF}_{\textrm{I}}}/dt$, of the LFBF$_{\textrm{I}}$. The shock speed decreases from $\sim$1287 km s$^{-1}$ to $\sim$405 km s$^{-1}$ in the distance range of $\sim$1.35--1.55 $R_\odot$ (see Appendix~\ref{appendixB}). These values are highly dependent on the FDR. The FDR is also a measure of non-radiality in the moving direction of the radio emission source. In our case, the FDR of the LFBF$_{\textrm{I}}$ rapidly decreases from $|-0.143|$ MHz s$^{-1}$ at 09:29:35~UT to $|-0.023|$ MHz s$^{-1}$ at 09:32:23~UT. Such fast dropping of the FDR during relatively short ($\sim$3 mins) lifetime of the type II burst can be caused by changing the moving direction of the radio source. We assume that at first the radio source moved radially from the Sun. The FDR values at the onset exceed $|-0.1|$ MHz s$^{-1}$ that is higher than typical FDR magnitudes of the type II bursts observed in decameter wavelengths~\citep{Aguilar-Rodriguez2005}. The subsequent fast decrease of the FDR implies that the radio source underwent a continual change in its direction. Apparently, at some instant, close to the onset of the radio emission, the radio source deviated from its initial course and kept propagating at an increasing angle to the radial direction. As a result, on the dynamic spectrum the LFBF$_{\textrm{I}}$ demonstrates smooth flattening. Therefore, the obtained shock speed values are accurate only at the radio emission onset.

\begin{figure}[!t]
   \centering
   \includegraphics[width=1.0\hsize]{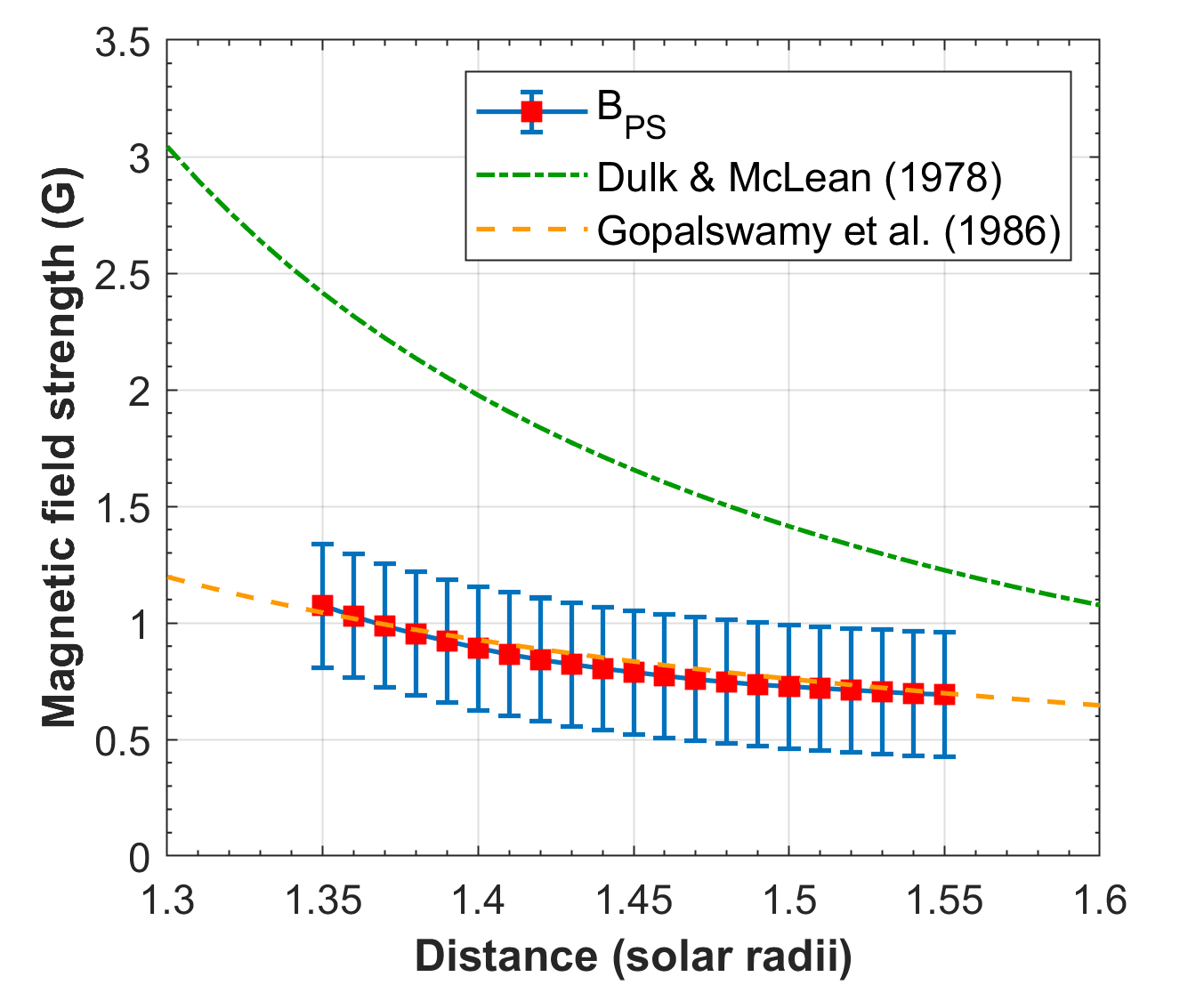}
   \caption{Magnetic-field strength at a lower part of the pseudo-streamer, $B_{\textrm{PS}}$, namely below its CP, as a function of distance from the photosphere ($1R_{\odot}$) calculated with the equation~(\ref{Eq1}). Main errors of $B_{\textrm{PS}}$ are due to a cloudy form of type II bursts on dynamic spectra, whereas instrumental errors are less 10\% of the former. The green and orange lines correspond to the coronal magnetic field models of \citet{Dulk1978} and \citet{Gopalswamy1986}, respectively.}
   \label{Figure4}
\end{figure}

There is one more strong argument supporting this scenario. The CME/shock speed in the low corona mimics the X-ray light curve during flaring events~\citep{Kumari2017}. The type II burst with spectral break coincides in time with the peak of the C1.6 solar flare. Indeed, the duration of the type II emission is centred at maximum of the X-ray flux in such a way that a half of the burst occurs before the flare's peak and a half (till the spectral break point) extends after it. It is unlikely that the shock speed drops rapidly around the maximum phase of the X-ray radiation~(Figure~\ref{Figure1}). It is more reasonably to suppose the shock speed in a close proximity to the peak of the X-ray flare profile changes very slowly within a narrow range or even takes on a constant value. Moreover, we performed a cross-check of the density jumps at the shock front for the cases of varying and constant shock speed~(see Appendix~\ref{appendixC}). There is almost an exact match between observational and theoretical density jumps for the constant shock speed equal to $\sim$1287 km s$^{-1}$. This value is regarded correct taking into account all the evidences.

Finally, the magnetic-field strength within the PS's base was determined using equation~(\ref{Eq1}). Figure~\ref{Figure4} shows magnetic fields in 1.08--0.69 G for the heliocentric distances of $\sim$1.35--1.55 $R_\odot$, found from the analysis of the type II burst with spectral break. The values are consistent well with the \citet{Gopalswamy1986} model and below the magnitudes given by the \citet{Dulk1978} model. The \citet{Gopalswamy1986} empirical model describes the ambient coronal magnetic field above mild active regions. It is in accordance with our event, since the flare and the ensuing CME took place in the unipolar AR 10572 of moderate activity. The feasibility of such a case was noted by~\citet{Stanislavsky2015}.

\subsection{``Fractured'' type II burst with spectral bump}

\begin{figure}[!t]
\begin{center}
\includegraphics[width=1.0\hsize]{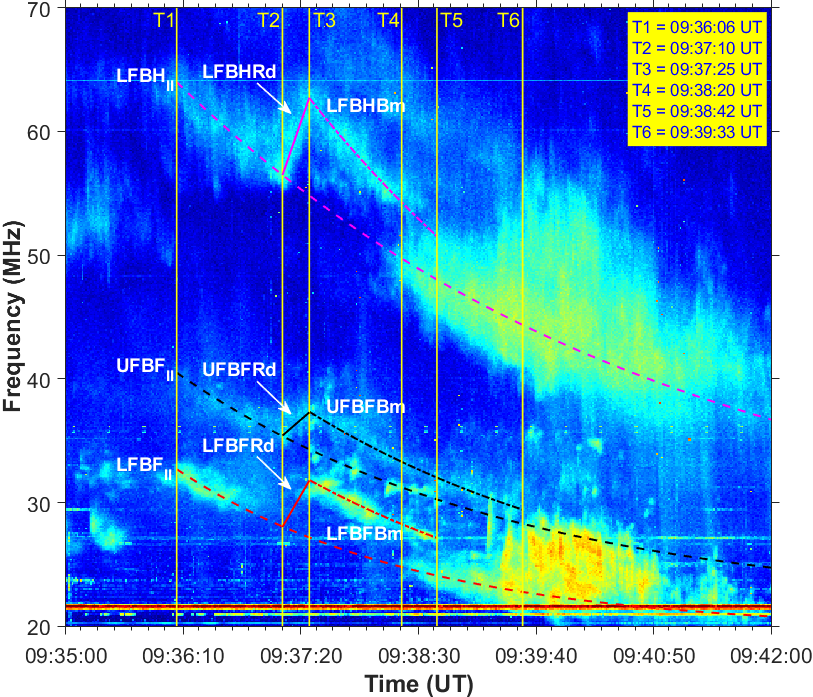}
\caption{The NDA dynamic spectrum of the type II radio burst with spectral bump. Lower and upper frequency bands of the fundamental component are labeled as LFBF$_{\textrm{II}}$ and UFBF$_{\textrm{II}}$, whereas the lower frequency band of the harmonic component is marked as LFBH$_{\textrm{II}}$. The bump structure presenting in LFBF$_{\textrm{II}}$, UFBF$_{\textrm{II}}$ and LFBH$_{\textrm{II}}$ are designated as LFBFRd-LFBFBm, UFBFRd-UFBFBm and LFBHRd-LFBHBm, respectively. Fitting with the exponential (LFBF$_{\textrm{II}}$, UFBF$_{\textrm{II}}$, LFBH$_{\textrm{II}}$, LFBFBm, UFBFBm, LFBHBm) and linear (LFBFRd, UFBFRd, LFBHRd) functions is performed. The yellow vertical lines indicate time instants T1-T6.}
\label{Figure5}
\end{center}
\end{figure}

\begin{figure}[!h]
   \centering
   \includegraphics[width=.99\hsize]{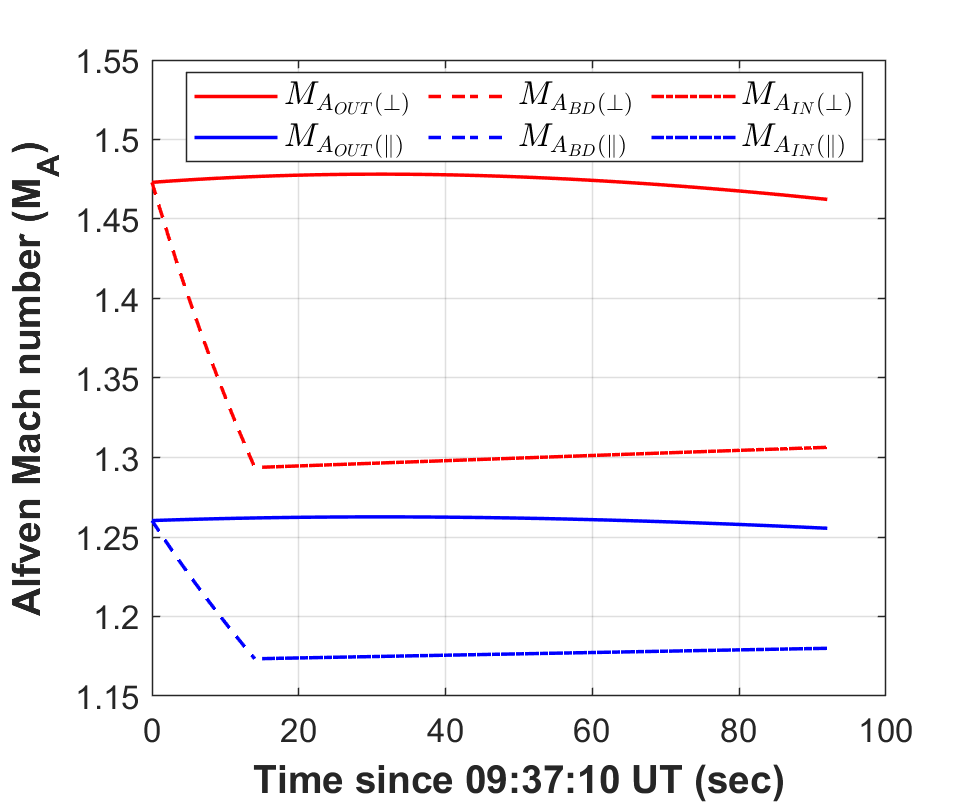}
   \caption{The Alfv\'{e}n Mach number values, $M_{A}$, determined by using equations~(\ref{Eq5}) and (\ref{Eq6}).}
   \label{Figure6}
\end{figure}

\begin{figure*}[!t]
\begin{center}
    \includegraphics[width=0.9\textwidth]{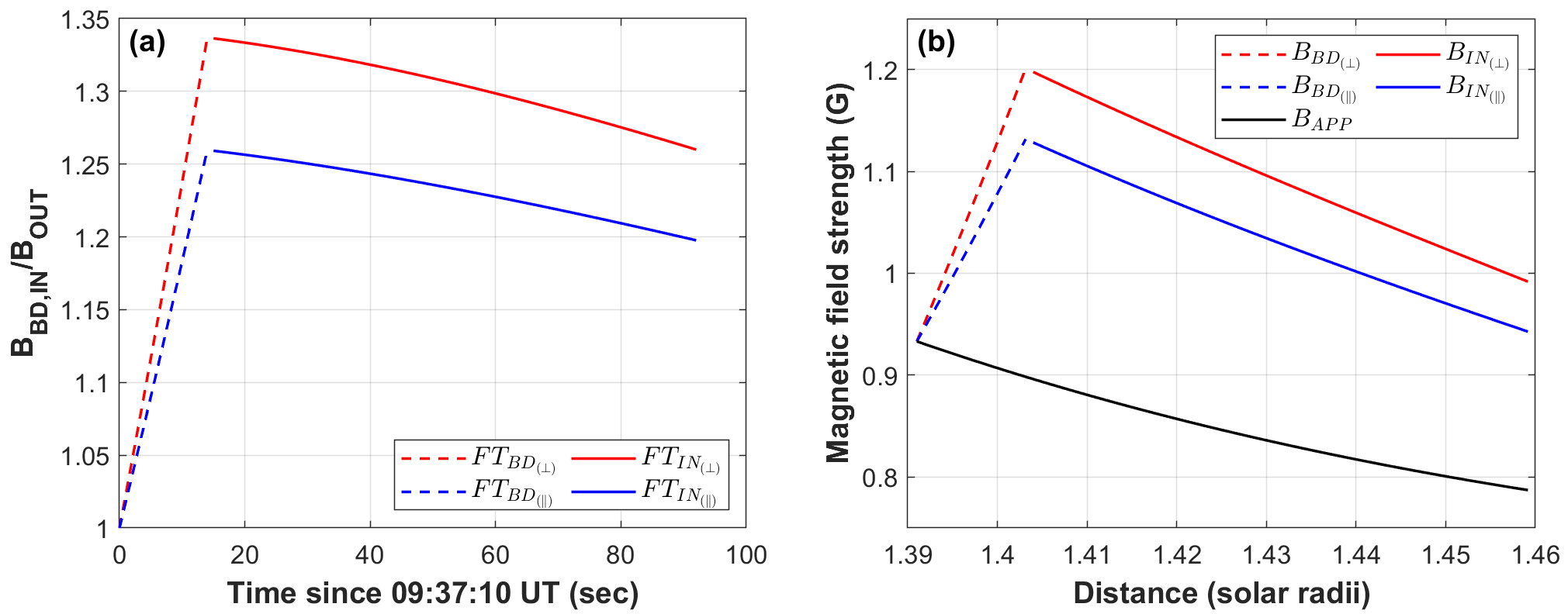}
    \caption{(a) Magnetic-field strength ratios in the FT's boundary and inside. (b) The absolute magnetic-field values in the FT's boundary and inside. The dark solid curve represents the approximation of the magnetic field values within the distance range of 1.391--1.459 $R_\odot$ obtained by the examination of the first ``fractured'' type II burst.}
\label{Figure7}
\end{center}
\end{figure*}

Figure~\ref{Figure5} displays the NDA dynamic spectrum with the second ``fractured'' type II burst. It started at 09:36:06 UT (T1) and, likewise its forerunner, has a harmonic structure with the band-splitting. The lower and upper frequency bands of the fundamental (F) component are labeled as LFBF$_{\textrm{II}}$ and UFBF$_{\textrm{II}}$; the lower frequency band of the harmonic (H) component is named as LFBH$_{\textrm{II}}$. The main feature of this burst begins at 09:37:10 UT (T2), when the frequency drift of all bands becomes reversal. These parts, designated as LFBFRd, UFBFRd, and LFBHRd, have positive drift rates (solid lines in Figure~\ref{Figure5}). The reversal radio emission has gone up in frequency for 15 seconds. After this all bands reversed their drifts again, thus making zigzag frequency variations. Since 09:37:25~UT (T3) the bands continued their gradual decrease, but well above the frequency drift pathways obtained from the exponential fitting of the LFBF$_{\textrm{II}}$, UFBF$_{\textrm{II}}$, and LFBH$_{\textrm{II}}$ (dashed lines in Figure~\ref{Figure5}). Such a spectral feature is identified as a type II spectral bump. Accordingly, the radio emissions lasting from 09:37:25~UT (T3) are dubbed as LFBFBm, UFBFBm, and LFBHBm (dash-dotted lines in Figure~\ref{Figure5}). The LFBFBm and LFBHBm become discontinuous at 09:38:42~UT (T5). Shortly before, at 09:38:20~UT (T4), the LFBF$_{\textrm{II}}$ and UFBF$_{\textrm{II}}$ showed up. So, the radio emissions of the LFBF$_{\textrm{II}}$ and UFBF$_{\textrm{II}}$ as well as LFBFBm and UFBFBm, coexisted during 22 seconds. The UFBFBm was terminated at 09:39:33~UT (T6). After that instant, the (F)-(H) components long for approximately 2.5 minutes like broadband, structureless stripes.

Based on the imaging study the CME/shock--FT interaction was identified. In this scenario, the radio emission source was located at the shock flank. The shock wave path laid through the FT, which was denser than the ambient coronal plasma. Drastic jumps of the electron density resulted in strong variations of the plasma frequency observed as a spectral bump in the type II dynamic spectrum. The band-splitting was present during the lifetime of the type II burst. Therefore, the magnetic-field strength values in the ambient plasma as well as in the FT can be calculated in the time interval from 09:37:10~UT (T2) to 09:38:42~UT (T5). However, the travel direction of the shock wave cannot be established reliably with the available observational data. Hence, the shock speed at the flank -- the key parameter for computing the magnetic-field strength -- becomes uncertain. To cancel the uncertain parameter and determine the magnetic-field strength magnitudes in the FT, we have introduced and applied the original approach described below.

The band-splitting of the fundamental component of the type II burst with spectral bump structurally consists of three parts. They are resulted from the shock wave passage through the region in the corona with enhanced electron density. The UFBFRd--LFBFRd and UFBFBm--LFBFBm pairs relate to the shock moving over boundary ($\textit{BD}$) and inside ($\textit{IN}$) of the FT, respectively. The UFBF$_{\textrm{II}}$--LFBF$_{\textrm{II}}$ pair corresponds to the shock traveling outside ($\textit{OUT}$) the FT, in the corona with monotonically decreasing the electron density. Thus, the density jumps across the shock front, $X$, can be calculated for each part as
\begin{equation*}
\begin{gathered}
X_{BD} = (f_{\textrm{UFBFRd}}/f_{\textrm{LFBFRd}})^2,\\
X_{IN} = (f_{\textrm{UFBFBm}}/f_{\textrm{LFBFBm}})^2,\\
X_{OUT} = (f_{\textrm{UFBF}_{\textrm{II}}}/f_{\textrm{LFBF}_{\textrm{II}}})^2.
\end{gathered}
\end{equation*}
The frequency drift values were obtained from the NDA spectrogram (see Table~\ref{Table2} in Appendix~\ref{appendixA}).

Knowing the density jumps, the Alfv\'{e}n Mach numbers, $M_{A}$, can be determined. Two extreme cases of the perpendicular and parallel shocks to the magnetic field are considered,  since the shock-to-magnetic-field orientation is unknown. In the case of the perpendicular shock, the Alfv\'{e}n Mach numbers are:
\begin{equation}
\begin{gathered}
M_{A_{BD}\perp} = \sqrt{X_{BD}(X_{BD}+5)/2(4-X_{BD})},\\
M_{A_{IN}\perp} = \sqrt{X_{IN}(X_{IN}+5)/2(4-X_{IN})},\\
M_{A_{OUT}\perp} = \sqrt{X_{OUT}(X_{OUT}+5)/2(4-X_{OUT})}.
\end{gathered}\label{Eq5}
\end{equation}
In the case of the parallel shock, the Alfv\'{e}n Mach numbers become:
\begin{equation}
\begin{gathered}
M_{A_{BD}\parallel}=\sqrt{X_{BD}},\\
M_{A_{IN}\parallel}=\sqrt{X_{IN}},\\
M_{A_{OUT}\parallel}=\sqrt{X_{OUT}}.
\end{gathered}\label{Eq6}
\end{equation}
Figure~\ref{Figure6} shows the computed Alfv\'{e}n Mach numbers in the FT's boundary and inside, as well as outside for the cases of perpendicular and parallel shocks.

Consequently, the magnetic-field strength magnitudes in the FT's boundary, $B_{\textrm{BD}}$, inside, $B_{\textrm{IN}}$, and outside, $B_{\textrm{OUT}}$, the FT for both cases can be expressed as
\begin{equation}
\begin{gathered}
B_{BD(\perp,\parallel)} = 5.1\cdot10^{-5}\cdot f_{\textrm{LFBFRd}}\cdot V_{S_{F}}/M_{A_{BD(\perp,\parallel)}},\\
B_{IN(\perp,\parallel)} = 5.1\cdot10^{-5}\cdot f_{\textrm{LFBFBm}}\cdot V_{S_{F}}/M_{A_{IN(\perp,\parallel)}},\\
B_{OUT(\perp,\parallel)} = 5.1\cdot10^{-5}\cdot f_{\textrm{LFBF}_{\textrm{II}}}\cdot V_{S_{F}}/M_{A_{OUT(\perp,\parallel)}},\\
\end{gathered}
\label{Eq7}
\end{equation}
where $V_{S_{F}}$ is the speed of the shock flank, i.e. unknown quantity here. To avoid this unknown quantity from the calculations, we determine the magnetic field values in the FT's boundary and inside relative to the corresponding values outside:
\begin{equation}
\begin{gathered}
\frac{B_{BD(\perp,\parallel)}}{B_{OUT(\perp,\parallel)}}=\frac{f_{\textrm{LFBFRd}}}{f_{\textrm{LFBF}_{\textrm{II}}}}\cdot\frac{M_{A_{OUT(\perp,\parallel)}}}{M_{A_{BD(\perp,\parallel)}}},\\
\frac{B_{IN(\perp,\parallel)}}{B_{OUT(\perp,\parallel)}}=\frac{f_{\textrm{LFBFBm}}}{f_{\textrm{LFBF}_{\textrm{II}}}}\cdot\frac{M_{A_{OUT(\perp,\parallel)}}}{M_{A_{IN(\perp,\parallel)}}}.
\end{gathered}
\label{Eq8}
\end{equation}

To evaluate the ratios (see Figure~\ref{Figure7}(a)), we exploited the results obtained from the analysis of the first ``fractured'' type II burst, assuming a steady behavior of the magnetic field and electron density in the vicinity of the AR 10572. The radiation frequencies $f_{\textrm{LFBF}_{\textrm{II}}}$ within the T2--T5 time span correspond to the radio-emitting source moving in the ambient corona as if there was no FT on its way. The frequencies within that span provide electron density values, $N_e$, as $f_{\textrm{LFBF}_{\textrm{II}}}\propto \sqrt{N_e(r)}$, where $r$ is the distance in solar radii. Applying the partial Saito's model of the electron density, the heliocentric distances were found. In other words, they are coronal heights at which the radio-emitting source traversed the FT. In this way, the $(B_{\textrm{BD}}/B_{\textrm{OUT}})$ and $(B_{\textrm{IN}}/B_{\textrm{OUT}})$ ratios became functions of $r$. It can be reasonably treated that the magnetic field variations in the ambient plasma, i.e., outside the FT, are permanent close to the AR 10572. This variation is ascertained from the examination of the type II burst with spectral break. Thus, we get the radial dependence of the magnetic field values $B_{\textrm{APP}}(r)$.

Eventually, the absolute values of the magnetic-field strength in the FT's boundary and inside the FT were determined with the expressions $(B_{\textrm{BD}}/B_{\textrm{OUT}})\cdot B_{\textrm{APP}}$ and $(B_{\textrm{IN}}/B_{\textrm{OUT}})\cdot B_{\textrm{APP}}$. Figure~\ref{Figure7}(b) shows the magnetic-field strength in the the FT's structure. Two extreme cases corresponding to parallel and perpendicular shock waves are indicated by blue and red colors. The magnetic-field strength in the boundary of the FT increases from 0.933 G at 1.391~$R_\odot$ to 1.128 G (blue dashed line) and to 1.2 G (red dashed line) at 1.403~$R_\odot$. Inside the FT the magnetic field magnitudes decrease within the range of 1.128--0.943 G (blue solid line) and 1.2--0.99 G (red solid line) in the interval of the heliocentric distances of 1.403--1.459 $R_\odot$. Note, the actual magnetic field values are contained in the region among blue and red lines.

\section{Discussion and conclusions}
\label{Sect:Discussion}

\begin{figure}[!h]
   \centering
   \includegraphics[width=.6\hsize]{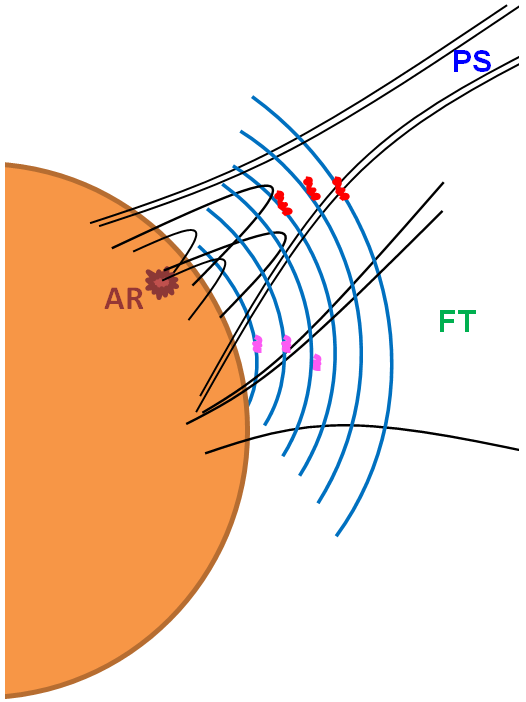}
   \caption{Sketch demonstrating the physical origin of both type II radio bursts. The magnetic field lines of the pseudo-streamer and magnetic flux tube as well as the locations of shock wave fronts are delineated by black and blue curves, respectively. The radio-emitting regions relating to the type II bursts with spectral break and bump are indicated by red and magenta thick segments, correspondingly.}
   \label{Figure8}
\end{figure}

We have reported spectral observations of two ``fractured'' type II solar bursts. The joint analysis of spectral and imaging measurements has allowed us to unequivocally identify coronal structures in the solar atmosphere responsible for the bursts' features in dynamic spectra. It was established that both ``fractured'' bursts were produced by the same shock caused by the CME after C1.6 class solar flare from the AR 10572. The AR was located under the PS, so the rising CME/shock penetrated into the PS from the photosphere that resulted in occurrence of the spectral break in the first type II burst. The spectral bump feature in the second type II burst emerged from the sideward passing of the same shock wave through the nearby FT. The last finding should be regarded as a novelty, since this designates FTs as coronal structures influencing the propagation of shock waves in the solar corona. So far, the type II spectral bump has been attributed to the propagation of a shock wave across a streamer stalk from one side to another~\citep{Feng2012,Feng2013}. The emergence mechanism of the type II radio bursts with break and bump is sketched in Figure~\ref{Figure8}.

In this way for the first time, we have obtained the magnetic-field strength in the PS and FT by the examination of ``fractured'' type II bursts. Foremost, the magnetic field inside a FT has been determined from radio observations. The strength values correspond to the ones known in literature~\citep{Woolsey2014,Pinto2017}. These results cope with a relevant task in the solar radio astronomy -- measurements of magnetic field in the coronal structures in a routine way. There are several semi-empirical extrapolation models exploiting photospheric field maps to estimate the coronal field typically up to 2.5 $R_\odot$~\citep{Wang1992,Abbo2015}. These models cannot reconstruct the exact magnetic configuration of the corona. They only provide an approximation of the global magnetic topology. In this connection, the type II bursts, showing the band-splitting and distinctive fractured shapes, can be as probing signals to get the magnetic-field strength in different coronal structures acting on the solar wind.

The present scientific research combines a novelty, relevance, and attractiveness. This study has strong impact on the analysis of the ``fractured'' radio bursts and contribute to the current studies of solar wind sources to which PSs and FTs belong. We encourage further investigations of ``fractured'' type II solar bursts as an alternative up-to-date method providing valuable information mainly on the magnetic-field strength in solar wind structures.

\acknowledgments

A.K., M.K., M.B. acknowledge support from the project RVO:67985815 and Grants: 20-09922J, 20-07908S of the Czech Science Foundation. A.S. thanks the Polish National Agency for Academic Exchange (NAWA PPN/ULM/2019/1/00087/DEC/1) for their support. We thank the radio monitoring service at LESIA (Observatoire de Paris) to provide value-added data that have been used for this study. The authors are also thankful to the SOHO, NOAA, RSTN, and AIP teams for their instrument maintenance and open data access. SOHO is a project of international cooperation between ESA and NASA.

\appendix

\section{Fitting the frequency drift of bursts}
\label{appendixA}

We estimate the frequency drift for both type II bursts by using exponential and linear functions as an approximation~\citep{Cunha-Silva2015}. To do this, we adjust exponential and linear fitting curves until they matched best visually with the emission lanes on the NDA dynamic spectrum. The fitting parameters for the first and second ``fractured'' type II bursts are contained in Table~\ref{Table1} and Table~\ref{Table2}, respectively.

\begin{table}[h!]
  \begin{center}
    \caption{Fitting parameters for approximation of the frequency drift of the type II burst with spectral break. All the frequencies are expressed in MHz.}
    \label{Table1}
    \begin{tabular}{c|c|c|c|c|c}
     \hline
      \multicolumn{6}{c} {$f = A\cdot \textrm{exp}(-Bt)+C$} \\
      \hline
      \textbf{ } & \textbf{A} & \textbf{B} & \textbf{C} & \textbf{t, s} & \textbf{t$_{st}$, UT}\\
      \hline
      \textbf{$f_{\textrm{LFBF}_{\textrm{I}}}$} & 13 & 0.011 & 18 & 0:168 & 09:29:35\\
      \textbf{$f_{\textrm{UFBF}_{\textrm{I}}}$} & 19 & 0.01 & 25 & 0:168 & 09:29:35\\
      \textbf{$f_{\textrm{LFBH}_{\textrm{I}}}$} & 31 & 0.009 & 31 & 0:168 & 09:29:35\\
      \textbf{$f_{\textrm{UFBH}_{\textrm{I}}}$} & 40 & 0.011 & 47 & 0:168 & 09:29:35\\
      \hline
      \textbf{$f_{\textrm{UFBFBr}}$} & 6 & 0.05 & 22.5 & 0:35 & 09:32:23\\
      \textbf{$f_{\textrm{UFBHBr}}$} & 14 & 0.05 & 39 & 0:35 & 09:32:23\\
      \hline
    \end{tabular}
  \end{center}
\end{table}

\begin{table}[h!]
 \begin{center}
    \caption{Fitting parameters for approximation of the frequency drift of the type II burst with spectral bump. All the frequencies are expressed in MHz.}
    \label{Table2}
    \begin{tabular}{c|c|c|c|c|c}
      \hline
      \multicolumn{6}{c} {$f = A\cdot \textrm{exp}(-Bt)+C$} \\
      \hline
      \textbf{ } & \textbf{A} & \textbf{B} & \textbf{C} & \textbf{t, s} & \textbf{t$_{st}$, UT}\\
      \hline
      \textbf{$f_{\textrm{LFBF}_{\textrm{II}}}$} & 13 & 0.007 & 19.7 & 0:354 & 09:36:06\\
      \textbf{$f_{\textrm{UFBF}_{\textrm{II}}}$} & 19 & 0.005 & 21.5 & 0:354 & 09:36:06\\
      \textbf{$f_{\textrm{LFBH}_{\textrm{II}}}$} & 39 & 0.0034 & 25 & 0:354 & 09:36:06\\
      \hline
      \textbf{$f_{\textrm{LFBFBm}}$} & 14.8 & 0.005 & 17 & 0:77 & 09:37:25\\
      \textbf{$f_{\textrm{UFBFBm}}$} & 16.8 & 0.005 & 20.5 & 0:128 & 09:37:25\\
      \textbf{$f_{\textrm{LFBHBm}}$} & 40.9 & 0.0042 & 21.8 & 0:77 & 09:37:25\\
      \hline
      \multicolumn{6}{c} {$f = Dt+F$} \\
      \hline
      \textbf{ } & \multicolumn{2}{c|}{\textbf{D}} & \textbf{F} & \textbf{t, s} & \textbf{t$_{st}$, UT}\\
      \hline
      \textbf{$f_{\textrm{LFBFRd}}$} &  \multicolumn{2}{c|}{0.269} & 28.02 & 0:15 & 09:37:10\\
      \textbf{$f_{\textrm{UFBFRd}}$} &  \multicolumn{2}{c|}{0.134} & 35.41 & 0:15 & 09:37:10\\
      \textbf{$f_{\textrm{LFBHRd}}$} &  \multicolumn{2}{c|}{0.447} & 56.47 & 0:15 & 09:37:10\\
      \hline
    \end{tabular}
  \end{center}
\end{table}

\section{Shock wave velocity estimates}
\label{appendixB}

The plasma frequency, $f_p$, varies with the electron plasma density, $N_e$, according to the following relation: $f_p[\textrm{MHz}] = 9\cdot10^{-3}\sqrt{N_e[\textrm{cm}^{-3}]}$. The frequency drift rate of the LFBF$_{\textrm{I}}$ of the type II burst with spectral break, $df_{\textrm{LFBF}_{\textrm{I}}}/dt$, can be used to determine the shock wave speed at its nose, $V_{S_{N}}$, using the following relation:
\begin{equation}
     V_{S_{N}} = \frac{df_{\textrm{LFBF}}}{dt}\frac{2N_e}{f}\left(\frac{dN_e}{dR}\right)^{-1}\frac{1}{\cos\theta},
     \label{Eq2}
\end{equation}
where $\theta$ is the deviation angle from the radial direction. In the present case we assumed $\theta = 0$. The frequency drift of the LFBF$_{\textrm{I}}$ was estimated directly from the NDA dynamic spectrum (see Appendix~\ref{appendixA}). Its time derivative -- the frequency drift rate of the LFBF$_{\textrm{I}}$ -- takes the next form: $\frac{df_{\textrm{LFBF}_{\textrm{I}}}}{dt} [\textrm{MHz}~\textrm{s}^{-1}] = -0.143\cdot \textrm{exp}(-0.011t[\textrm{s}])$.

\begin{figure}
   \centering
   \includegraphics[width=.85\hsize]{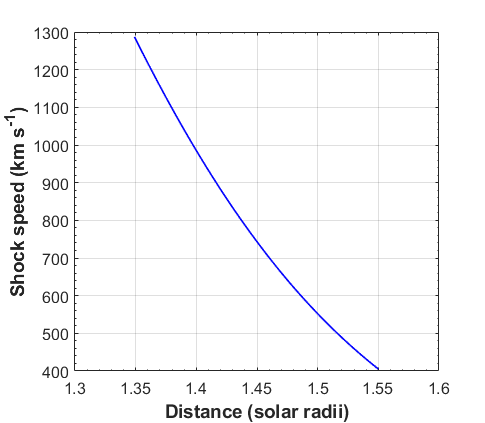}
   \caption{Shock wave speed at its nose as a function of distance from the photosphere (1R$_{\odot}$) computed by using equation~(\ref{Eq2}).}
   \label{Figure9}
\end{figure}

To obtain the shock wave velocity, the correct model of electron density should be applied. Since the electron density decreases radially with the distance from the photosphere, $N_e = N_e(R)$, it is possible to determine the density and/or plasma frequency at a specific distance. It was estimated that the apex of the semi-circular arch, containing magnetically closed region of the PS, locates at $\sim$1.5 $R_\odot$. It is considered as an upper limit in distances where the penetration of the PS by the shock could take place. On the other hand, the spectral break frequency in the LFBF$_{\textrm{I}}$ is 20.05 MHz. This puts constraints on the selection of density model, since the above-mentioned values should conform to each other. Thus, we used the partial Saito density model with a multiplier factor of 0.4~\citep{Saito1977}. According to it, the plasma frequency of 20.05 MHz corresponds to the height of 1.55~$R_\odot$. The velocity of the shock wave is calculated and presented in Figure~\ref{Figure9}.

\section{Cross-comparison of electron density jumps}
\label{appendixC}

\begin{figure}[!b]
   \centering
   \includegraphics[width=.99\hsize]{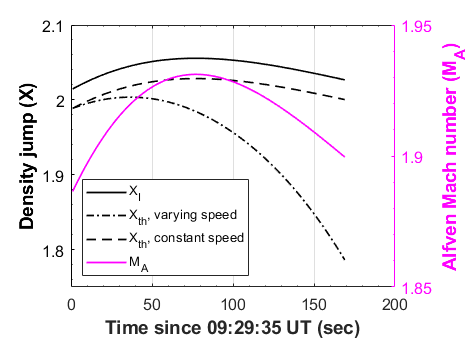}
   \caption{Observational, $X_{\textrm{I}}$, and theoretical, $X_{th}$, density jumps: $X_{\textrm{I}}$ calculated from the the band-splitting of the fundamental component of type II burst with spectral break from the NDA spectrogram (black solid curve); $X_{th}$ found by using the equation~(\ref{Eq4}) for the cases of varying (black dash-dotted curve) and constant (black dashed curve) speed of the shock wave. The Alfv\'{e}n Mach numbers, $M_{A}$, determined with the equation~(\ref{Eq3}) are represented by a solid magenta curve.}
   \label{Figure10}
\end{figure}

The density jump across the shock front, $X$, characterizes the downstream-to-upstream shock compression ratio. In the given study, we can obtain the density jump from the band-splitting of the type II burst with spectral break as $X_{\textrm{I}} = (f_{\textrm{UFBF}_{\textrm{I}}}/f_{\textrm{LFBF}_{\textrm{I}}})^2$. Under the quasi-perpendicular shock-to-magnetic-field approximation and the low plasma-to-magnetic pressure ratio, the Alfv\'{e}n Mach number is related to the density jump $X_{\textrm{I}}$ as
\begin{equation}
     M_{A} = \sqrt{X_{\textrm{I}}(X_{\textrm{I}}+5)/2(4-X_{\textrm{I}})}.
     \label{Eq3}
\end{equation}
Figure~\ref{Figure10} shows the computed density jump, $X_{\textrm{I}}$, and the Alfv\'{e}n Mach numbers, $M_{A}$. The frequency drift values of the UFBF$_{\textrm{I}}$ and LFBF$_{\textrm{I}}$ were taken from the NDA spectrogram (see Table~\ref{Table1} in Appendix~\ref{appendixA}).

The density jump, $X_{\textrm{I}}$, obtained from the observational data can be compared with its theoretical value, $X_{th}$, to ensure their consistence. To do this, we refer to the theory of the interstellar shocks that is applicable to the solar coronal shocks~\citep{Draine1993}. The formula of the density jump for the quasi-perpendicular shock wave has the following form:
\begin{equation}
     X_{th} = 2(\gamma+1)/(D + [D^2 + 4(\gamma+1)(2-\gamma)M_{A}^{-2}]^{1/2}),
     \label{Eq4}
\end{equation}
where $D\approx(\gamma-1)+(2C_{s}^2/V_{S_{N}}^2 + \gamma/M_{A}^2)$. The values of the sound speed $C_{s}\approx 128$ km s$^{-1}$~\citep{Gopalswamy2012} and the adiabatic index $\gamma = 5/3$ \citep{Vrsnak2002} correspond to the solar corona in the distance range 1.3--1.6 $R_{\odot}$. Finally, the $X_{th}$ magnitudes can be obtained using the numerical values of $C_{s}$, $\gamma$, $M_{A}$ (from equation~(\ref{Eq3})), and $V_{S_{N}}$ (from equation~(\ref{Eq2})).

Therefore, we have calculated the density jump, $X_{th}$, for two cases, namely for varying and constant speed of the shock wave (for an explanation see the ``Fractured'' type II burst with spectral break'' subsection). In the first case, the complete $V_{S_{N}}$ value set was used (black dash-dotted curve in Figure~\ref{Figure10}). In the second case, the $V_{S_{N}}$ magnitude was a constant of about 1287 km s$^{-1}$ corresponding to the first element of the set (black dashed curve in Figure~\ref{Figure10}). There is a visible discrepancy between $X_{\textrm{I}}$ and $X_{th}$ (``varying speed'') values, whereas the $X_{\textrm{I}}$  and $X_{th}$ (``constant speed'') curves coincide well.

\end{document}